# Optical frequency comb Fourier transform cavity ring-down spectroscopy

ROMAIN DUBROEUCQ[1] AND LUCILE RUTKOWSKI[1,*]

[1]*Univ Rennes, CNRS, IPR (Institut de Physique de Rennes)-UMR 6251, F-35000 Rennes, France*
*\*lucile.rutkowski@univ-rennes1.fr*

**Abstract:** We demonstrate broadband and sensitive cavity ring-down spectroscopy using a near infrared frequency comb and a time-resolved Fourier transform spectrometer. The cavity decays are measured simultaneously and spectrally sorted, leading to purely exponential decays for each spectral element. The absorption spectra of atmospheric water and carbon dioxide are retrieved and demonstrate the high frequency resolution and absorption precision of the technique. The experimental apparatus, the measurement concept and the data treatment are described. The technique benefits from the advantages of cavity ring-down spectroscopy, i.e. the retrieved absorption does not depend on the cavity parameters, opening up for high accuracy absorption spectroscopy entirely calibration-free.

## 1. Introduction

Cavity ring-down spectroscopy (CRDS) is a well-established and robust spectroscopic technique. It enables highly accurate absorption measurements of gaseous samples in various environments [1]. It relies on the use of an optical cavity injected with a resonant radiation and consequently measuring the decay rate of the radiation through the cavity once the light source has been shutoff [2–4]. The decay rate variation is directly linked to the intra-cavity absorption, which makes the absorption spectra inherently calibration-free, i.e. not requiring the knowledge of the physical separation or the reflectivity of the cavity mirrors. When combined with continuous wave laser diodes [5], CRDS provides high-resolution molecular absorption spectra, and has been used for a wide variety of applications including (but not limited to) atmospheric field measurement [6], thermometry [7], saturated spectroscopy [8–10], breath analysis [11], absorption metrology [12–14], laboratory astrophysics [15,16], etc.

One of the main remaining challenges of CRDS is to achieve broadband spectroscopy in limited acquisition times. Broadly tunable continuous wave lasers are commercially available but suffer from slow tuning speed, even if mode-hop-free operation is achieved. This leads to long measurement times and can induce bias as the sample environment may vary during the acquisition. Therefore, there is a strong interest in obtaining spectra with the same resolution in a multiplex fashion, using a broadband light source. In a first attempt [7], the parallel spectral acquisition was performed using time-resolved Fourier transform spectroscopy (FTS). This acquisition technique was initially developed to monitor biological samples [8] and was later applied to gaseous molecular samples [19]. It relied on a step-scanning FTS where the time-dependent event is triggered at each optical path difference (OPD) step of the interferometer, generating a time-resolved signal once the entire path difference has been scanned. This way, it was possible to reach resolutions sufficient to resolve atomic and molecular transitions. Engeln and Meijer used the same approach to record the CRDS decays of a picosecond laser and retrieved the doublet transitions of $O_2$ near 765 nm [17]. However, the total acquisition time (4 hours) and the sensitivity ($2.7\times10^{-7}$ cm$^{-1}$ for a cavity finesse around 4000 and a resolution of 0.4 cm$^{-1}$) were still prohibiting many applications.

Optical frequency combs have proven versatile to perform cavity enhanced spectroscopy [20]. Typically generated by femtosecond laser sources, they provide a broadband and high-resolution source that allows for efficient coupling in optical cavities [21,22]. These unique characteristics have led to the development of many cavity enhanced spectroscopy approaches, most of them requiring knowledge of the cavity parameters to be fully quantitative. In 2007, Thorpe et al. published the first attempt of comb-based CRDS [23]. The setup combined the CRDS cavity with a dispersion spectrometer and a camera detection, leading to an acquisition time of 1.4 ms, and a resolution of 0.8 cm$^{-1}$, limited by the resolution power of the dispersion grating (no sensitivity was extracted from the final spectrum). Very recently, Lisak *et al.* demonstrated comb-based CRDS using a dual-comb detection approach [24]. This approach also relies on the measurement of time-resolved interferograms, but thanks to the high acquisition speed of a single interferogram (5 µs), several interferograms are recorded during a single cavity decay to yield the decay time-dependence. This setup demonstrated a sensitivity of $3\times10^{-8}$ cm$^{-1}$ for an acquisition time of 1s and a cavity finesse around 18'000.

Here, we demonstrate the first implementation of CRDS based on an optical frequency comb source and time-resolved FTS detection (FT-CRDS). The frequency comb is tightly locked to the cavity and the light extinction is triggered at a constant OPD step in order to retrieve the spectrally resolved cavity decays. Using a balance detection in the FTS yield high signal-to-noise ratio decay signals that are rearranged in a time-resolved interferograms. The technique apparatus and measurement design are described in details, together with the different steps of the data processing. This approach allows measuring the entire spectral bandwidth transmitted by the cavity simultaneously with a high sensitivity and a short acquisition time. We demonstrate the potential for molecular spectroscopy by applying FT-CRDS to the detection of atmospheric species ($H_2O$ and $CO_2$) in an open-air cavity. Finally, we discuss the performances of the technique and its current limitations.

## 2. Experimental setup and data acquisition

The general experimental layout is depicted in Fig. 1(a). The optical frequency comb is generated by an amplified Er:fiber oscillator (Mode-Locked Technology, ModeHybrid) which delivers a power of 100 mW at a repetition rate of 100 MHz. The emitted light goes through an acousto-optic modulator (AOM, G&H, SFO1814-T-M080-0.4C2J-3-F2P-03) before reaching the ring-down cavity. The cavity is 75 cm long, i.e. every other comb mode is transmitted by the cavity, and it is opened to the laboratory air. Based on the specifications of the cavity mirrors (Layertec, coating A0612015, 99.88% reflectivity), and the cavity length, the ring-down time of the empty cavity is expected around $\tau_0 = 1.6$ µs. The comb is locked to the cavity using the Pound-Drever-Hall technique (PDH) to ensure a quasi-continuous transmission of the comb modes through the cavity. The PDH modulation is performed at 10 MHz using a fiber-coupled electro-optic modulator (EOM) inserted between the comb oscillator and amplifier. The PDH error signal is demodulated, filtered and amplified before being fed to two servos (New Focus, LB1005-S) controlling a slow piezo actuator and a second EOM, both placed inside the comb oscillator.

The transmitted light is analyzed using a Fourier transform spectrometer (FTS) designed after ref. [20] and depicted in Fig. 1(b). The OPD is scanned using two retro-reflectors mounted on a 30 cm-long translation stage (Thorlabs DDS300). The FTS is functioning in a fast-scanning mode, with the cart moving at a nearly constant speed of 5 mm/s during the entire measurement. The comb propagates on a twice folded optical path and the two output interferograms are monitored using two independent photodiodes with 13 MHz bandwidth (Thorlabs, PDA10CS2, PD1 and PD2 on Fig. 1(b)). The cart movement is monitored using a frequency stabilized HeNe laser (Thorlabs, HRS015B) which propagates in one of the interferometer arms. The HeNe light is coupled in fiber before reaching a splitter: 50% of the light is coupled to free-space and is reflected on one of the retro-reflectors before being coupled in fiber again and recombined with the remaining 50% that comes directly from the first splitter. The sinewave interferogram it produces is used as a reference signal for a direct digital generator (DDG, Berkeley Nucleonics Corp 745T) which generates a boxcar signal every time the cart moves of half a HeNe wavelength ($\lambda_{HeNe}$). This boxcar signal is used to control the AOM and start each cavity decay, and its duration is set to 12 µs (approx. $7\tau_0$). The same boxcar signal is used to trigger the acquisition card (National Instrument, PCI5922,

256 Mbits/channel). The PD1 and PD2 signals are acquired simultaneously, and the acquisition card records 180 points at a 15 Ms/s sampling rate and a 16 bits resolution for each decay. In total, 400'000 decays are acquired, yielding total OPD of 50.6 cm for the two comb interferograms.

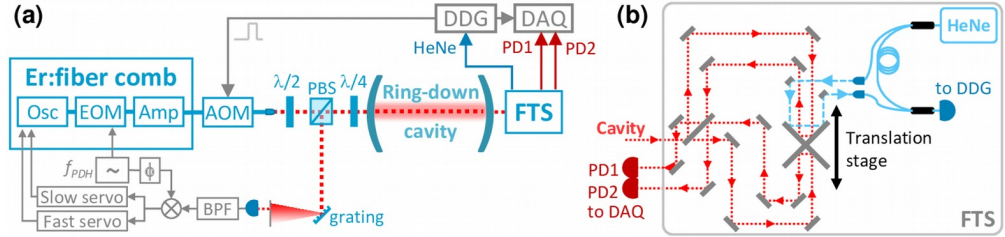

Fig. 1. Experimental setup. (a) General layout: EOM, electro-optic modulator; AOM, acousto-optic modulator; λ/2, half-wave plate; PBS, polarization beam splitter; λ/4, quarter-wave plate; PDH, Pound-Drever-Hall locking electronics; FTS, Fourier transform spectrometer; DDG, digital delay generator; DAQ, acquisition card. (b) FTS design: the two FTS outputs are detected on independent photodiodes (PD1 and PD2) sent to the DAQ, and the optical path difference (OPD) is calibrated using a HeNe laser. Note that the OPD varies twice faster for the comb interferogram than for the HeNe interferogram for the same cart spatial displacement.

The raw data recorded on PD1 is shown in Fig. 2. It contains the decays measured at the different OPD steps. The envelope of the decays follows the intensity of the two-sided interferogram of the comb and the center burst is visible at 12 s, corresponding to the zero OPD point. Part of the signal is zoomed in Fig. 2(b), revealing the successive decays. The red-circular markers show the acquired data. Each of these decays is a multi-exponential function, as it integrates the spectral variation of the ring-down time. It is noteworthy that the AOM chopping the comb also extinguishes the PDH error signal. The servo parameters have been optimized to minimize the impact of the chopping over the locking performances, yielding the direct cavity build-up once the comb light is injected again as depicted by the gray line in Fig. 2(b). The upper scales show the time scale for each decay. The FTS cart continuously moves during the decays, but this does not prevent the calibration of the interferogram at any given time after shutoff (e.g. considering every $n^{th}$ data point of each section) as the OPD step remains equal to $2\lambda_{HeNe}$. The fast scanning FTS operation brings a small phase shift on the interferogram as the time after shutoff increases, without noticeable impact on the spectrum.

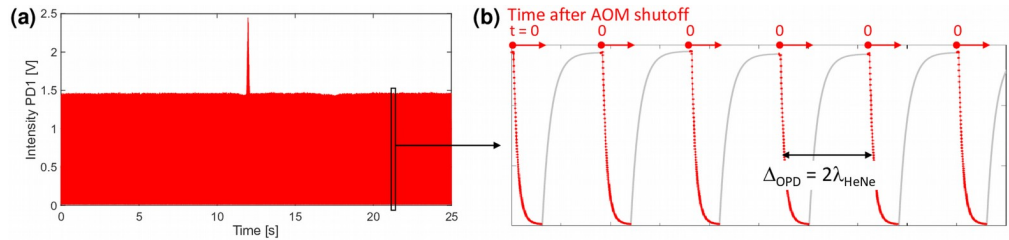

Fig. 2. (a) Raw data measured at one of the FTS outputs as the OPD is swept continuously. The center burst of the interferogram is visible around 12 s. (b) Enlargement of a small part of the signal, where the cavity decays occur at a constant OPD step. The red circle markers are the acquired data, the gray curve show the simulated cavity build-ups which are not acquired. The upper part on the panel shows the delay to the successive AOM triggers.

The raw data sets from photodiodes PD1 and PD2 are cut in sections after each AOM shutoff and re-organized in 3D interferograms, such as the one displayed in Fig. 3(a) showing the PD1 decays close to zero OPD. The flat intensity region at negative time values has been recorded before the comb light is shutoff and is consequently removed from the dataset. Fig. 3(b) plots the PD1 and PD2 interferograms retrieved at t = 0 µs, which are affected by a strong intensity noise. This is caused by the locking constraints but also by the small timing jitter of the DAQ (internal clock yields a 67 ns uncertainty on the acquisition start after the DDG trigger).

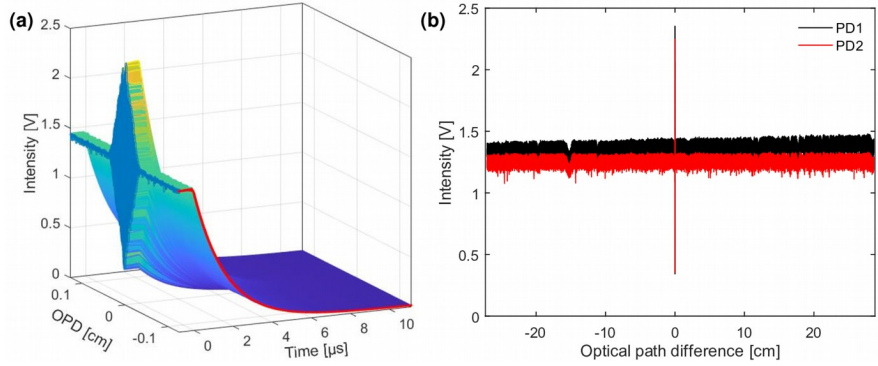

Fig. 3. (a) 3D interferogram (enlargement close to zero OPD) retrieved from the signal recorded on PD1. The first 12 points of each section are measured before the AOM shutoff. (b) Interferograms retrieved at t = 0 µs from PD1 (black line) and PD2 (red line).

## 3. Data processing

The first step of data processing involves cancelling the intensity noise using a numerical balancing approach similar to the autobalancing approach described in previous works [20]. For each time delay, the ratio of the PD1 and PD2 interferograms is calculated and fitted with a linear function which gives the intensity discrepancy between the two acquisitions. The balanced interferogram is then calculated subtracting the weighted PD2 interferogram from the PD1 and the same procedure is repeated for all time delays. Assessing the gain function for the successive interferograms as the decay occurs allows compensating for the detectors non-linearity as well and increase the robustness of data treatment. The result of the balancing procedure is shown in Fig. 4. Panel (a) shows the same part of the 3D interferogram as Fig. 3(a), where the DC component at each time delay has been cancelled and the decays are now only visible in the burst amplitude. Panel (b) contains the interferogram obtained at t = 0 s, with a signal-to-noise ratio improved by a factor >20 compared to the interferograms shown in Fig. 3(b). The remaining noise is limited by the two photodiodes PD1,2.

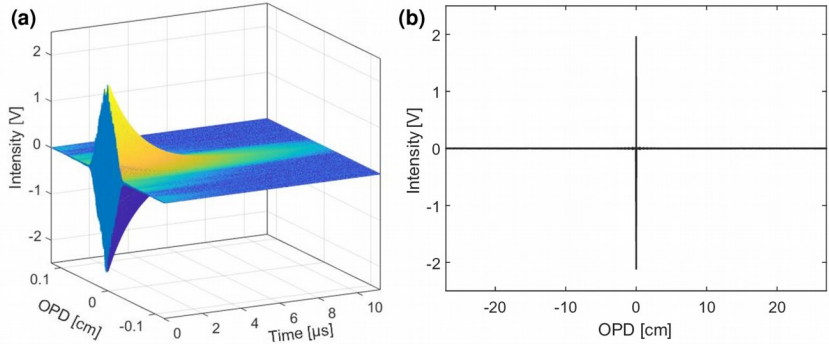

Fig. 4. Balanced 3D interferogram: (a) enlargement of the decays close to zero OPD, (b) full balanced interferogram obtained at t = 0 µs.

The spectral distribution of the decays is calculated next by taking the absolute value of the FFT of the interferograms at every time delay. This yields the 3D spectrum shown in Fig. 5(a). The amplitude of the decays is now proportional to the intensity of the comb transmitted by the cavity, and some $CO_2$ transitions are already visible on the spectrum at t = 0 s. Retrieving the ring-down spectrum implies to consider the decays at each spectral element. As an example, the normalized cavity decay at 6325 cm$^{-1}$ is shown in Fig. 5(b) together with a fit of the exponential decay function: $I(t) = I_0 e^{-t/\tau_{rd}} + y_0$, where $t$ is the decay time and the fitted parameters are $I_0$, the source intensity spectrum; $\tau_{rd}$, the ring-down time; and $y_0$, a detection offset. The fitted ring-down time is equal to 1.59 µs and the residuum is shown in the lower part of the panel. It does not exhibit any structure, showing that the FT-CRDS approach allows to successfully untangle the multi-ringdown decays in separate spectral elements. The standard deviation of the residuum over the full decay is equal to $1.2\times10^{-3}$. The same fit is repeated on the cavity decays at every spectral element to get the spectral variation of the ring-down time.

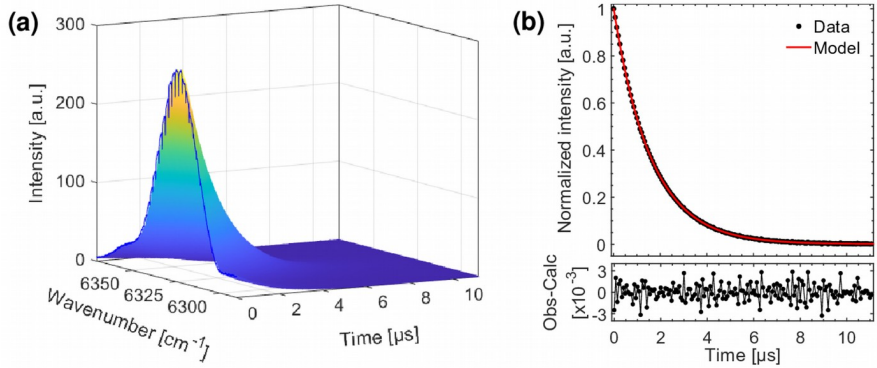

Fig. 5. (a) 3D spectrum obtained after fast Fourier transformation of the interferograms at each distinct decay time. (c) Typical measured decay measured at 6325 cm$^{-1}$ (upper window, black circles) plotted with the fitted exponential decay (upper window, red line) and the residuum of the fit in the lower window.

## 4. Atmospheric detection

The ring-down time at a specific wavenumber $\nu$ is related to the empty cavity ring-down time $\tau_0$ and to the absorption of the intra-cavity molecular sample $\alpha$ via the relationship $[c\tau_{rd}(\nu)]^{-1} = [c\tau_0(\nu)]^{-1} + \alpha(\nu)$ where $c$ is the speed of light in vacuum. Fig. 6(a) shows the spectral variation of $1/[c\tau_{rd}]$ over the entire range of the comb transmitted through the cavity. The wavenumber axis is calibrated using the HeNe wavelength value and one of the molecular line center frequencies from the HITRAN database [25]. This spectrum can be modelled as a slowly varying baseline summed to the absorption spectra of $H_2O$ and $CO_2$, which are the two main atmospheric species absorbing in this range. The baseline retrieved from the fit is shown in Fig. 6(a) and it is subtracted from the spectrum to isolate the molecular contributions. This last spectrum is shown in the upper panel of Fig. 6(b). The panel below shows the synthetic spectra of $H_2O$ and $CO_2$ (respectively the blue and red curves, inversed for clarity) calculated using Voigt absorption profiles and the spectroscopic parameters available in the HITRAN database. The only fitted parameters were the concentrations of both species and their resulting values are [$H_2O$] = 1.48% and [$CO_2$] = 0.100%. These values are consistent with expectations in the ambient air of a closed laboratory. The noise equivalent absorption is found to be $1.5\times10^{-8}$ cm$^{-1}$ for a resolution of 0.02 cm$^{-1}$ (600 MHz), value given by the standard deviation of the residuum on the spectral range in the shaded area. The acquisition time was 25.3 s for a single spectrum and the spectrum contains 2000 spectral elements. The residuum exhibits some remaining distortions at the positions of a few transitions in the center of the spectral range, but they appear to be uncorrelated to the linestrength of the transition. Further investigations have shown the critical need to achieve the FTS scan at constant speed, and this structure may result from the small variations of the cart speed when acquiring the raw data.

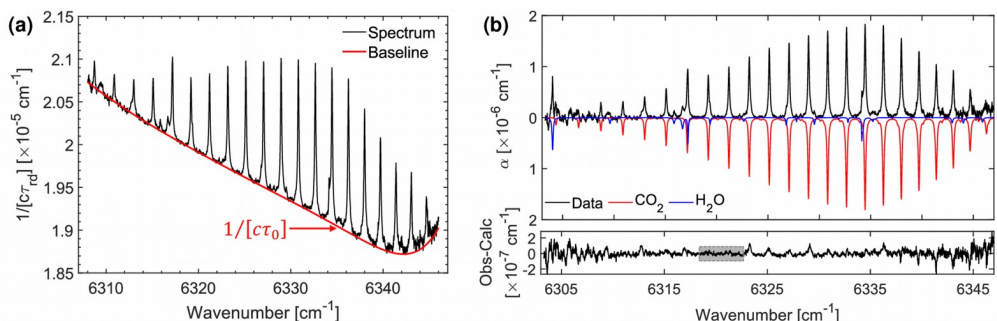

Fig. 6. (a) Absorption spectrum obtained from the ring-down times (black curve) and the fitted baseline (red curve). (b) Measured absorption spectrum with the baseline subtracted (upper panel, black line) plotted together with the simulated spectra of $CO_2$ (red curve) and $H_2O$ (blue curve) calculated using the HTRAN parameters and Voigt absorption profiles and whose concentrations have been linearly fitted.

## 5. Discussion

In its current version, the FT-CRDS setup allows measuring calibration free absorption spectra and relies on the frequency calibration of the Fourier transform spectrometer. Once the ring-down spectrum is obtained, the analysis is straightforward: as any CRDS spectrum, it can be modelled by a linear summation of the absorption of the sample species with a slowly varying baseline. The spectral bandwidth is inherently limited to the transmission of the comb through the cavity, and hence is impacted by the performances of the locking electronics and by the cavity finesse and dispersion. On the other hand, the spectral resolution and frequency precision are not fundamentally limited by the nominal resolution of the FTS, as the technique is fully compatible with the sub-nominal method [26]. The latter only requires matching the nominal resolution to the mode spacing of the comb transmitted through the cavity and yield spectra benefiting from the comb mode precision and resolution. This development will require stabilizing the cavity length (e.g. following the procedure shown in [27]). The cavity is currently open to air and its length is left free-running, which prevents the application of the sub-nominal method but also prevents averaging successive measurements.

In addition to averaging, achieving a higher sensitivity should be feasible by increasing the cavity finesse. However, implementing a cavity with a longer ring-down time will affect strongly the locking performance, probably to the point where the lock will not hold during the shutoff duration (since it also suppresses the error signal). Therefore, reaching high finesses will require revising the locking approach, using e.g. a third laser referencing the cavity and the optical frequency comb. Nevertheless, the present system sensitivity is close to the value expected in a cavity enhanced configuration using the same cavity and considering a detection signal-to-noise ratio of 1000, which would be $\alpha_{min} = ¿ \, 8 \times 10^{-9}$ cm$^{-1}$. The FT-CRDS is yet immune to the line distortion caused by the cavity dispersion that affects the spectra produces by Fourier transform cavity enhanced spectroscopy based on frequency combs [28].

## 6. Conclusion

FT-CRDS based on an optical frequency comb source as been implemented for the first time and has proven a reliable method to record absorption signatures of small molecules. Time-resolved FTS was able to sort the cavity decays in the correct spectral element bins, opening up for precision spectroscopy of small molecules. First spectra were measured in laboratory air and have been modelled using the HITRAN database. The technique combines the advantages of CRDS based on continuous-wave lasers and the broad bandwidth of the comb source, with a short acquisition time given by the multiplex acquisition. Future works will include stabilizing the cavity length to allow an efficient averaging and measurements in a controlled environment to assess the accuracy performances of the system.

**Funding.** This work is supported by the ANR (project CECoSA, ANR-19-CE30-0038-01), the Physics National Institute (INP CNRS), the scientific council of Région Bretagne and Rennes métropole.

**Acknowledgments.** The authors acknowledge Aleksandra Foltynowicz for the loan of the cavity mirrors and Oliver Heckl for proof-reading the manuscript.

**Data availability.** Data underlying the results presented in this paper are not publicly available at this time but may be obtained from the authors upon reasonable request.